# Multilayer Network Planning - A Practical Perspective

**Achim Autenrieth**
*ADVA Optical Networking SE, Martinsried / Munich, Germany*
*aautenrieth@advaoptical.com*

**Abstract:** The paper presents a pragmatic and practical multilayer network planning approach based on a candidate lightpath auxiliary graph model. The paper discusses, how this approach can be applied to offline network planning as well as dynamic planning and provisioning of services.
**OCIS codes:** (060.0060) Fiber optics and optical communications; (060.4254) Networks, combinatorial network design

## 1. Introduction

This paper presents a flexible multilayer network planning approach for optical transport networks from a practical perspective. The goal is to achieve an optimized balance between electrical switching and grooming and transparent optical bypass. The approach presented in this paper describes the planning practice used in several technology evaluation and research studies, e.g. [1, 2], as well as in commercial tenders. To address such diverse use cases, the planning approach must be flexible and pragmatic, as it must be able to adapt to and cope with different requirements. The multilayer network planning approach is based on a greedy multilayer heuristic using an auxiliary graph to model feasible candidate lightpaths and to solve the virtual topology design problem. In the next sections, the steps of the planning workflow as implemented in multilayer network planning tool will be presented.

## 2. Multilayer Network Planning Process

Figure 1 shows the workflow and individual tasks of the multilayer planning process. The multilayer graph model is using four main graphs as shown in the lower part of Figure 1: the demand graph, the fiber graph, a virtual topology graph and an auxiliary graph which represents the candidate (optically feasible) lightpaths. The output of the planning process contains the detailed routing and grooming information, a bill of material, and key performance metrics and statistics to evaluate and compare the quality of the solution. Graphical visualization of the output such as the shown wavelength allocation table support the analysis of the solution.

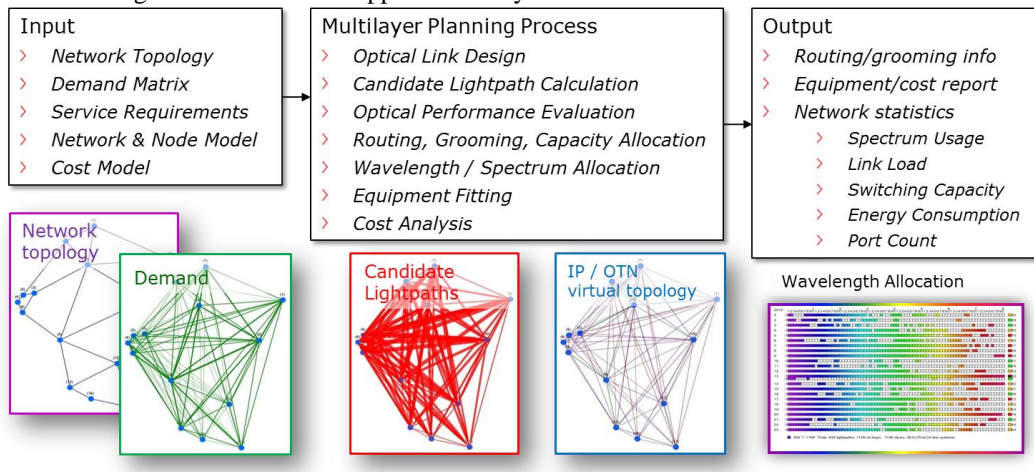

Fig. 1. Laser

The auxiliary graph model facilitates the optical performance evaluation as well as the efficient routing and grooming assignment using optimized optical bypass. The multilayer grooming heuristic was introduced in [1,2]. In the next sections, the steps and tasks of the planning workflow will be presented in detail.

*Customer requirements and input data*
The primary input to the planning tool is the network topology and the demand matrix. The network topology contains node sites and fiber links. The demand matrix may contains different services types like Ethernet, IP/MPLS and OTN services (odu0, odu1, odu2). For each demand, the service type and the bitrate or number of services in case of TDM services is given in the demand matrix. The planning tool must also be able to address different service requirements and technology constraints.

- Services with different protection and restoration requirements, e.g. some services require optical protection, some require optical restoration, and some both.
- Explicit routing of services
- Mix of fixed, direction-less, color-less ROADMs
- fixed or flexible WDM grid
- Supported transponder types, line rates, modulation formats, and slot widths
- Internal equipment constraints

*Optical Link Design*
The first step in the planning process is the optical link design. The design goal is the maximization of transparent reach. Ideally, all node-pairs can be transparently reached using the shortest path, or the shortest pair of disjoint paths for protection. The maximum reach is impacted by the distribution of span loss, the number of ROADMs in the path, linear and non-linear impairments. Optical network design must guarantee error-free operation of all installed and provisioned lightpaths over the whole lifetime, temperature range and all configurations (channel load etc.). The optical link design is done for end-of-life configurations that contain aging margins from the system vendor as well as margins given by the operator e.g. for span repair.

*Candidate Lightpath Calculation*
The multilayer planning approach is based on an auxiliary graph model using a so-called candidate lightpath (CLP) graph. For every node-pair a set of distinct routes is calculated: the k-shortest fiber paths, and the shortest link or node disjoint paths for optical protection. If the optical layer supports restoration, the path calculation is repeated for every expected failure event, e.g. all single link failures. Each resulting path is represented by an edge in the candidate lightpath graph and contains the fiber route as attribute. So, for each node pair, several parallel edges are present in the CLP graph.

*Optical Path Performance Evaluation*
For all edges in the CLP graph, an optical path performance evaluation is done to validate the feasibility of the lightpath. Evaluation of calculated end-to-end candidate lightpaths can be done using cascading formulas for different optical impairments followed by comparison against vector of threshold values. The threshold vector depends on the transponder type, modulation format, channel, grid type & spectral bandwidth, neighboring channels, and used or supported Forward Error Correction (FEC) and includes all required margins like span repair, ROADM pass-through, or ageing margins.

Unfeasible candidate lightpaths which exceed a threshold metric are removed from the candidate lightpath graph. The result is a graph with feasible candidate lightpaths. In case of a network with multiple line-rates or software-defined transponder with variable modulation formats and slot widths, the candidate lightpath may only be feasible for a limited set of bitrates and modulation schemes. For each edge in the candidate lightpath layer the parameters resulting in a feasible transmission are stored as edge attributes.

*Virtual Topology Design*
A candidate lightpath is selected as a grooming link if the potentially carried traffic is high enough to justify the bypass lightpath. This relates to a traffic grooming or virtual topology design problem. The goal of the grooming heuristics is to minimize the number of line interfaces by installing bypass lightpaths.

In the presented planning approach, the candidate lightpath layer is used to model candidate grooming links. For each candidate lightpath, the potential grooming load is calculated by computing $k$ candidate grooming paths for all demands on the candidate lightpath layer, and summing up the potential grooming load of each candidate grooming link / lightpath. The candidate lightpath is selected as grooming link and added to the virtual topology, if the grooming load exceeds a threshold. In our studies we use the grooming threshold as a sensitivity parameter and vary the values between 10% and 100% of the link capacity with the selected modulation format / line rate. If a candidate bypass lightpath does not satisfy the grooming load threshold requirement, it is removed from the list of candidate lightpaths (grooming links). This is repeated until all candidate lightpaths are processed (either selected or deleted from the list of candidate lightpaths). At low network load, the grooming heuristic yields a nearly opaque network with only very few transparent lightpaths with few bypass hops. At high network load, the resulting network is getting more and more transparent, i.e. the number of virtual links is increasing. If the demand between every node pair is larger than the grooming threshold, the resulting virtual network is fully meshed and converges to the transparent network case.

*Demand Routing, Grooming and Capacity Allocation*

When the virtual topology is defined, the demands are routed on the grooming links, and the required capacity is allocated on each virtual link. If not enough link capacity is available, a lightpath is installed and provisioned in the virtual link, thus increasing the capacity of the grooming link.

*Lightpath Installation and Wavelength / Spectrum Allocation*

When a lightpath is installed, the wavelength or spectrum must be allocated depending on the used WDM grid. Many wavelength and spectrum assignment algorithms are published in literature, the most common being the First Fit wavelength assignment. In case of a flexible WDM grid and modulation format with different slot widths, spectrum allocation algorithms should be preferred which reduce spectrum fragmentation and minimize or avoid the need for spectrum defragmentation.

The planning tool should also support parallel fibers or automatic fiber overbuild of bottleneck links if the capacity of a fiber is exhausted. For most operators, fiber overbuild is preferred to a longer fiber route.

*Equipment fitting / Bill of Material Calculation*

The final step is the equipment fitting and generation of the bill of material to calculate the cost of the network, and to evaluate the quality of the resulting network in terms of cost, energy efficiency, available capacity, and spectral resource usage.

## 3. Application for Dynamic Network Planning and Service Provisioning

In traditional optical network planning and operation, the network planning is separate from optical network operation and service provisioning with dedicated, independent tools. Optical network planning tools design and optimize the physical and photonic parameters of the network, which is then installed and deployed. Different tools and algorithms are used for network management and control plane during the operational phase of the network. Today the optical network is becoming more flexible and dynamic. SDN-based operation and SDN- or GMPLS-based multilayer interworking requires a dynamic service provisioning taking physical impairments into consideration. In the ACINO project [3], the dynamic multilayer service provisioning is extended to take application specific requirement into account using application-centric IP/optical network orchestration. For this, a centralized network orchestrator must be able to perform dynamic multilayer network planning and service (lightpath) provisioning. For dynamic service provisioning, the validated candidate lightpaths calculated in the multilayer network planning can be stored in a centralized network controller. The controller can expose and advertise the candidate lightpaths as abstract links in a virtual topology to an SDN controller. Alternatively, a path can be queried on demand using the PCE protocol or a RESTful API accessing an enhanced active stateful PCE [4]. Compared with an impairment-aware routing algorithms implemented directly in the PCE or centralized controller, this approach allows the inclusion of and compliance with vender and operator margins over the whole system lifetime.

## 4. Conclusion

In this paper we presented a multi-layer network planning approach for optical transport network. The proposed auxiliary graph model using a candidate lightpath layer allows a flexible and efficient modelling of fixed-grid and flex-grid optical networks with variable channels rates. The calculated candidate lightpaths are validated using cascading formulas against a set of threshold parameters. The same approach can be used for offline network planning as well as for dynamic online planning and service provisioning. In offline capacity planning, a candidate lightpath is selected as grooming link if the potential grooming load exceeds a configurable grooming threshold. This approach allow a guaranteed lightpath operation over the whole system lifetime.

*Parts of the presented work has received funding from the European Commission within the H2020 Research and Innovation Programme, under grant agreeement n.645127, project ACINO (www.acino.eu).*